\documentclass{emulateapj}
\usepackage{color}

\slugcomment{ApJ, accepted}
\shorttitle{MHS}
\shortauthors{Wiegelmann et al.}
\begin{document}
\title{Magneto-static modelling of the
mixed plasma Beta solar atmosphere
based on SUNRISE/IMaX data}
\author{\textsc
T.~Wiegelmann$^{1}$,
T.~Neukirch$^{2}$,
D.H.~Nickeler$^{3}$,
  S.K.~Solanki$^{1,6}$,
  V.~Mart\'\i nez Pillet$^{4}$,
  J.M.~Borrero$^{5}$}

\affil{$^{1}$Max-Planck-Institut f\"ur Sonnensystemforschung,
Justus-von-Liebig-Weg 3,
37077 G\"ottingen, Germany,
$^{2}$School of Mathematics and Statistics, University of St. Andrews,
St. Andrews KY16 9SS, United Kingdom,
$^{3}$Astronomical Institute, AV CR, Fricova 298,
25165 Ondrejov, Czech Republic,
$^{4}$National Solar Observatory, Sunspot, NM 88349, USA,
$^{5}$Kiepenheuer-Institut f\"ur Sonnenphysik, Sch\"oneckstr. 6,
79104 Freiburg, Germany,
$^{6}$School of Space Research, Kyung Hee University, Yongin,
Gyeonggi, 446-701, Korea.
}
\email{wiegelmann@mps.mpg.de}
\begin{abstract}
Our aim is to model the 3D magnetic field structure of
the upper solar atmosphere, including regions of non-negligible plasma beta.
We use high-resolution photospheric
magnetic field measurements from SUNRISE/IMaX as boundary condition
for a magneto-static magnetic field model. The high resolution of
IMaX allows us to resolve the interface region between photosphere
and corona, but modelling this region is challenging for the following
reasons.
While the coronal magnetic field is thought to be
 force-free (the Lorentz-force vanishes),
this is not the case in the mixed plasma $\beta$ environment
in the photosphere and lower chromosphere. In our model,
pressure gradients and gravity forces are taken self-consistently
into account and compensate the non-vanishing Lorentz-force.
Above a certain height (about 2 Mm) the non-magnetic forces become very weak
and consequently the magnetic field becomes almost force-free.
Here we apply a linear approach, where the electric current density
consists of a superposition of a field-line parallel current and a current
perpendicular to the Sun's gravity field. We illustrate the prospects and
limitations of this approach and give an outlook for an extension towards
a non-linear model.
\end{abstract}
\section{Introduction}
\begin{figure}
\includegraphics[width=0.5 \textwidth]{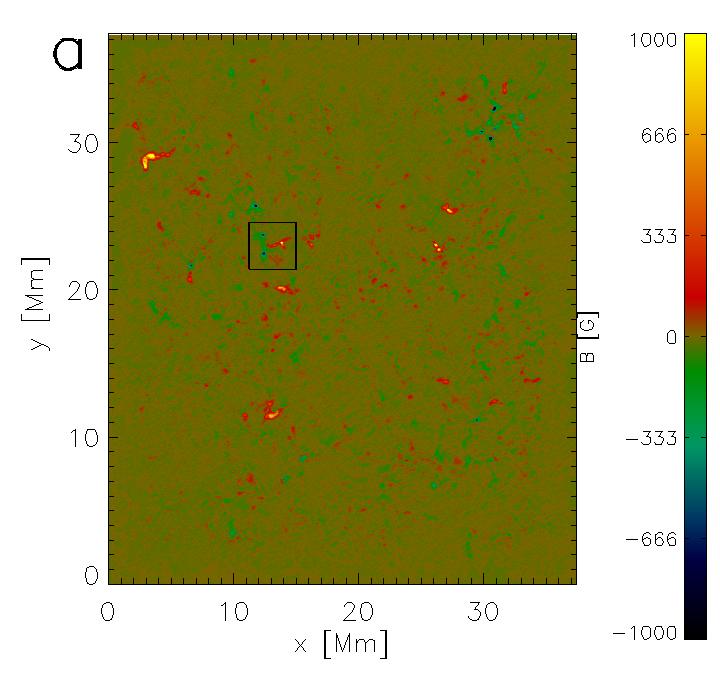}
\includegraphics[width=0.5 \textwidth, bb= 40 20 670 500, clip = true]
{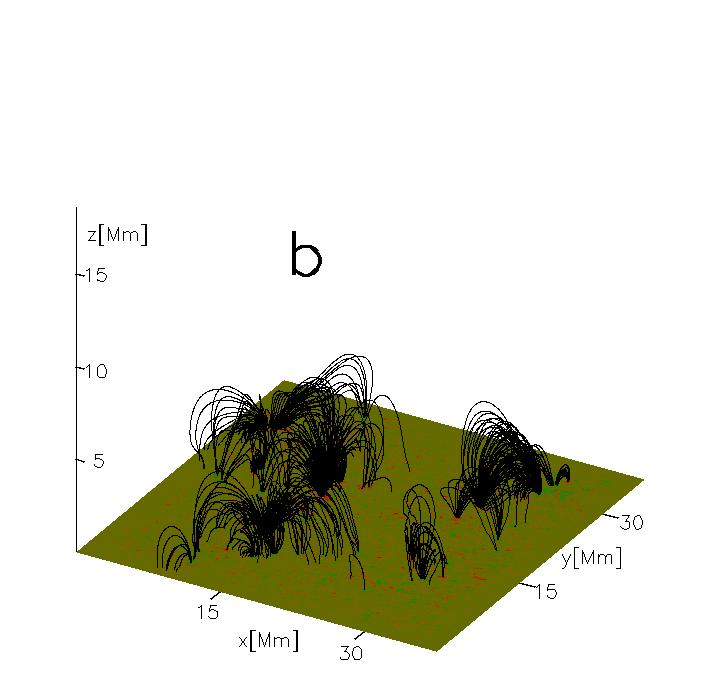}
\caption{Panel a: SUNRISE/IMax magnetogram of a quiet Sun area. The black
rectangular marks the region of interest.
Panel b: Sample field lines for a MHS-model.}
\label{sunrise_mag_full}
\end{figure}

While the corona, at least above active regions,
has a low plasma $\beta$ and is usually modelled by the assumption
of a vanishing Lorentz-force
\citep[see][for an overview of solar force-free fields]{2012LRSP....9....5W},
this is not true in the lower solar atmosphere
\citep[see][for a recent review on magnetic fields in the solar
atmosphere]{2014A&ARv..22...78W}. In the photosphere and lower chromosphere
low and high $\beta$ regions exist side by side and non-magnetic forces
have to be taken into account, to lowest order with a magneto-static model,
where the Lorentz-force is compensated by the gradient of the plasma
pressure and the gravity force.

The most accurate measurements of the
solar magnetic field are available in the photosphere. In active regions
the full magnetic vector can be measured accurately, e.g. with
SDO/HMI, whereas in quiet Sun regions only the line-of-sight or vertical
field is available with sufficient accuracy for a reliable extrapolation,
because in weak field regions there is too much uncertainty in the
transverse field components
\citep[Noise in the Stokes vector translate into an
uncertainty in the inferred values for the magnetic
field, see][]{2011A&A...527A..29B,2012A&A...547A..89B}.
These photospheric measurements are
extrapolated into the solar atmosphere under certain model assumptions, here
a magneto-static approach. The vertical resolution of the model scales
with the horizontal resolution of the photospheric measurements, e.g. about
1400 km for SOHO/MDI-magnetograms and 350 km for SDO/HMI. As the
non-force-free layer containing the photosphere and lower chromosphere is
rather thin (typically less than 2000 km), one can hardly resolve
magnetic structures here for models using SOHO/MDI- or SDO/HMI-magnetograms
as boundary condition. The high resolution
magnetograms from {\sc SUNRISE}/IMaX with a pixel size of only 40 km
allow now to model this layer vertically with about 50 points.

A special class of magneto-static solutions, which allow separable
solutions has been proposed by \cite{low91}. An advantage of this
approach is that the resulting equations are linear
\citep[for nonlinear cases, see][]{neukirch97} and
can be solved effectively by a Fourier transformation
or  a Green's function implementation
\citep[see][]{petrie:etal00}.
Separable and linear solutions have been found also
in spherical \citep{bogdan:etal86,neukirch95,al_sphere10}
as well as in cylindrical coordinates \citep[][]{neukirch09,al_cylinder10}.
Especially the solutions found in spherical coordinates have been used
for modelling the global magnetic field of the Sun
\citep[e.g.][]{bagenal:etal91,gibson:etal95,gibson:etal96,zhao:etal00,ruan:etal08}
and other stars \citep[e.g.][]{lanza08,lanza09}.

Usually these models require only the line-of-sight or vertical
photospheric magnetic field as boundary condition and the solutions
contain free parameters and/or free functions.
Nonlinear magneto-static solutions are more demanding numerically
and observationally, because they require photospheric
vector magnetograms as input
\citep[see][for a cartesian and spherical implementation,
respectively]{wiegelmann:etal06,wiegelmann:etal07}.
Within this work we apply the linear magneto-static solutions
proposed by \cite{low91} to a high-resolution magnetogram observed with
{\sc SUNRISE}/IMaX. We outline the paper as follows. In section
\ref{sec:basics} we briefly discuss the basic equations and
model assumptions. Section \ref{sec:data} describes the employed photospheric
magnetograms, which we use as boundary condition for our magneto-static
model in section \ref{sec:results}. In section
\ref{sec:outlook} we finally discuss the prospects and
limitations of this approach and give an outlook for a generalization
of the method towards a non-linear numerical approach.

\section{Basic equations}
\label{sec:basics}
We use the magneto-hydro-static equations
\begin{eqnarray}
{\bf j}\times{\bf B} & = &  \nabla P +\rho \nabla \Psi,
\label{forcebal}\\
\nabla \times {\bf B } & = &  \mu_0 {\bf j} , \label{ampere} \\
\nabla\cdot{\bf B}  & = &  0,    \label{solenoidal}
\end{eqnarray}
where ${\bf B}$ is the magnetic field,
${\bf j}$ the electric current density,
$P$ the plasma pressure,
$\rho$ the mass density, $\Psi$ the gravitational potential
 and $\mu_0$ the permeability of free space.
To find separable solutions for this set of equations, we
apply the following ansatz
for the electric current density \citep[see][for details]{low91}.
\begin{equation}
\nabla \times {\bf B } = \alpha_0 {\bf B } + f(z) \nabla B_z \times {\bf e_z},
\label{lin_mhs}
\end{equation}
where $\alpha_0$ is the force-free parameter and $f(z)$ is a free function,
which controls the non-magnetic forces.
The first part $\alpha_0 {\bf B }$ corresponds to a field-line-parallel
linear force-free current and the second term
$f(z) \nabla B_z \times {\bf e_z}$ defines a current perpendicular to
the gravitational force
 (in  the $z$-direction) or, in other words, parallel to the Sun's surface
$(x,y)$.
It is then possible to reduce the MHS equations to a single partial
differential equation
\citep[see e.g.][for a particularly simple formulation]{neukirch:etal99}
that can often be solved by separation of variables.
For convenience we use here \citep[as proposed in][]{low91}
\begin{equation}
f(z)= a \exp(-\kappa z),
\label{lin_mhs_2}
\end{equation}
with a free parameter $a$, which controls the non-magnetic forces in the
photosphere. Obviously, for the choice $a=0$, this approach reduces
to linear force-free fields. Above a certain height in the solar
atmosphere one expects that
the solution becomes approximately force-free, owing to the low
plasma $\beta$ in the solar corona. Consequently $f(z)$ has to decrease
with height and here we choose as a scale height
the distance of the upper chromosphere above the solar surface,
leading to $1/\kappa = 2 Mm$.
With $\kappa$ fixed, our MHS-solution contains two free parameters, $\alpha$
and $a$.
Let us remark that $\kappa$ in equation (\ref{lin_mhs_2})
controls the non-magnetic forces and should not be confused with
the scale height of the plasma pressure.

As boundary conditions we use the measured vertical magnetic field
$B_z(x,y,0)$ in the photosphere. We solve the equations by means of
a Fast-Fourier-Transform method similar to
the linear force-free model developed by
\cite{alissandrakis81}. Different from the linear force-free approach
is that the resulting Schr\"odinger equation for $B_z$ in the Fourier
space contains a Bessel function instead of an exponential function

One finds the following solution for pressure and density
\citep[see][for the derivation]{low91}

\begin{eqnarray}
P  & = & \; P_0(z) \; -\frac{1}{2 \mu_0} f(z) B_z^2, \label{pressure} \\
\rho & = & - \frac{1}{g} \frac{d P_0}{dz} 
+\frac{1}{\mu_0 g} \left[\frac{d f}{dz}\frac{B_z^2}{2}
+f \, ({\bf B } \cdot \nabla) B_z \right].
\end{eqnarray}
The first term in equation (\ref{pressure}) contains a 1D-solution
(in z-direction), which is independent of the magnetic field and
has to obey $\nabla P = -\rho \nabla \Psi$. The second term
is the disturbance of this 1D-pressure profile by the magnetic field. Here
pressure and density compensate the non-vanishing Lorentz-force.
This disturbance is negative (if $a > 0$), and obtains its highest absolute values
in regions of the highest vertical magnetic field strength $B_z$.
Because the total plasma pressure (sum of both terms) has to
be positive, we get the following in-equality for $P_0(z)$
\begin{equation}
P_0(z) >  a \cdot \exp(-k z) \cdot \frac{{\rm Max}(B_z)^2}{2 \mu_0}(z),
\label{P0}
\end{equation}

where $\frac{{\rm Max}(B_z)^2}{2 \mu_0}(z)$ is the maximum  at a given height
$z$.
As we will see later, this condition has severe consequences for an
application to data with strong locally enhanced magnetic fields in
the photosphere. To satisfy condition (\ref{P0}) in these regions, the
background pressure $P_0$ has to be so high that the plasma $\beta$ in
weak-field regions (and also on average) becomes unrealistically high,
see Fig. \ref{beta}.
Within this limitation, the choice of $P_0(z)$ has some freedom.
Our choice is given in section \ref{sec4.1}.

\section{Data}
\label{sec:data}
We apply our newly developed code to photospheric magnetic field
measurements taken with the balloon-borne {\sc SUNRISE}
solar observatory in June 2009.
For an overview of the  {\sc SUNRISE} mission
and scientific highlights of the first SUNRISE flight
see
\cite{solanki:etal10}, \cite{2011SoPh..268....1B},
\cite{2011SoPh..268..103B}, \cite{2011SoPh..268...35G}.
For a description of the IMaX instrument, we refer to
\cite{martinezpillet:etal11}. The photospheric magnetic field was
computed by inverting the IMaX measurements using the VFISV code as
described in \cite{borrero:etal11}.
The linear force-free extrapolation code, and the particular
case of an $\alpha=0$ potential field has been applied to data
from {\sc SUNRISE}/IMaX before for a single magnetogram by
\cite{wiegelmann:etal10} and to analyse a time series
by \cite{wiegelmann:etal13}.
\cite{chitta14} carried out non-linear force-free
extrapolations from IMaX magnetograms and added vertical
flows at low heights to simulate non-force-free effects
in the photosphere and chromosphere.
Here we apply our newly developed linear MHS-code to a snapshot
of the quiet Sun, observed with {\sc SUNRISE}/IMaX as well. We apply
our code first to the full field-of-view of IMaX, as shown in
Fig. \ref{sunrise_mag_full}
and in a subsequent step
we investigate a subfield (marked with a black rectangular in
Fig. \ref{sunrise_mag_full}) in more detail.
The data set used here was observed in a period of $1.616$ hours starting
at 00:00 UT on 2009 June 9th (image 220 from this series),
see \cite{martinezpillet:etal11}.

\section{Results}
\label{sec:results}
\begin{figure}
\includegraphics[width=0.5 \textwidth,height=4cm]{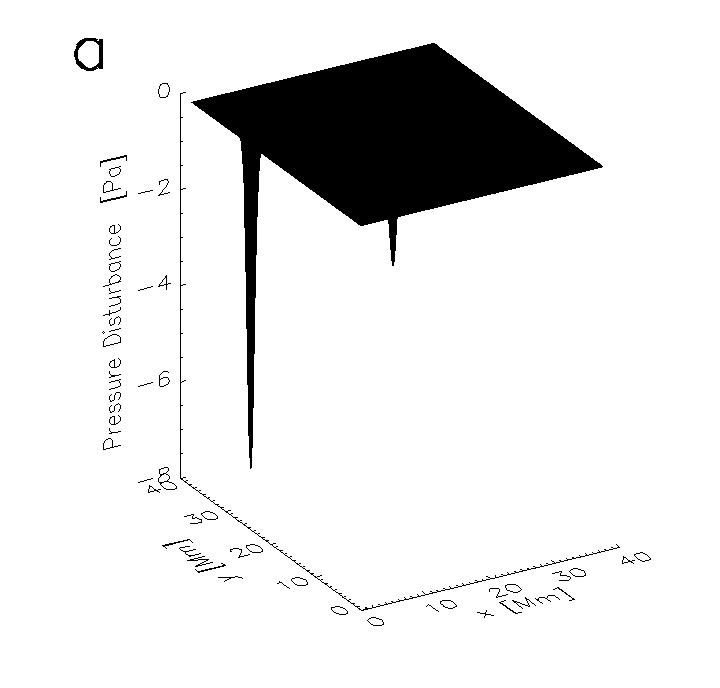}
\includegraphics[width=0.5 \textwidth,height=4cm]{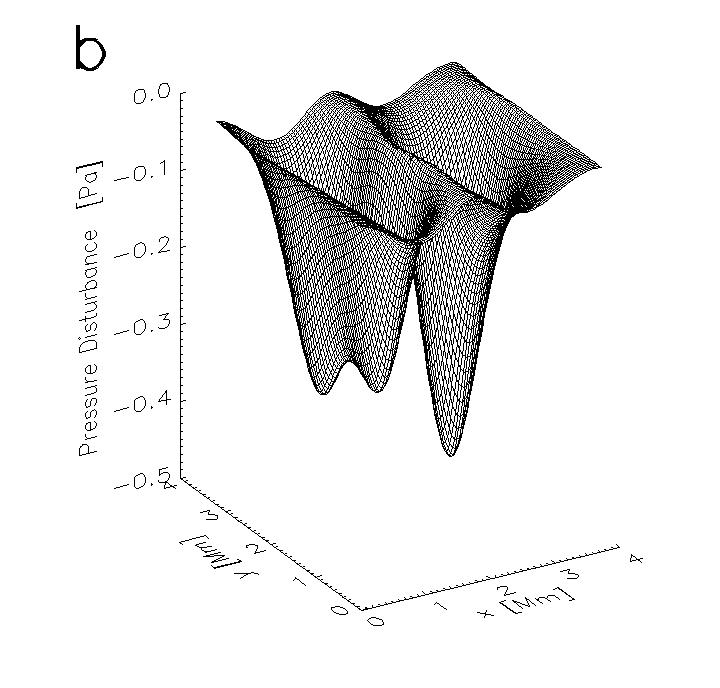}
\includegraphics[width=0.5 \textwidth,height=7cm]{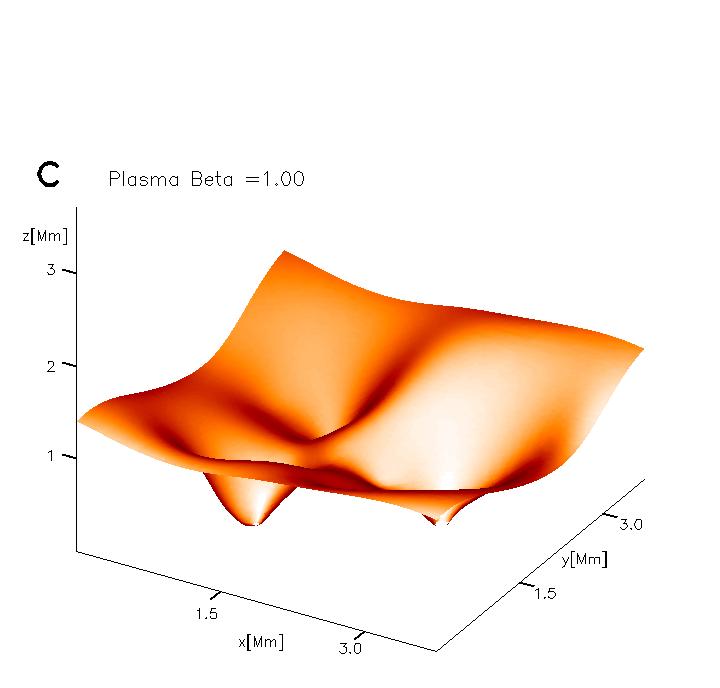}
\caption{The plasma pressure disturbance
$-\frac{1}{2 \mu_0} f(z) B_z^2$ at a height $z=1 Mm$ for the full IMaX
and the small FOV in panel a) and b), respectively.
Panel c) shows an equi-contour surface for $\beta=1$ in the
small FOV. }
\label{pic_pressure}
\end{figure}
\subsection{Application to the full IMaX-FOV.}
\label{sec4.1}
In our first computation we apply our model to the full
phase-diversity reconstructed IMaX magnetogram of
a quiet Sun region of $37 \times 37$ Mm, which has been resolved
by $936 \times 936$ pixels (pixel size on Sun 40 km),
see Fig. \ref{sunrise_mag_full},
As our main interest lies in the mixed
plasma $\beta$ regions of the photosphere and chromosphere, we extrapolate
up to a height of $z=4$ Mm  or 100 pixels. A few sample field lines for a magneto-static
solution with $\alpha=3.0$ and $a=0.5$ is shown in
in Fig. \ref{sunrise_mag_full} b).

\begin{figure}
\includegraphics[width=0.5 \textwidth]{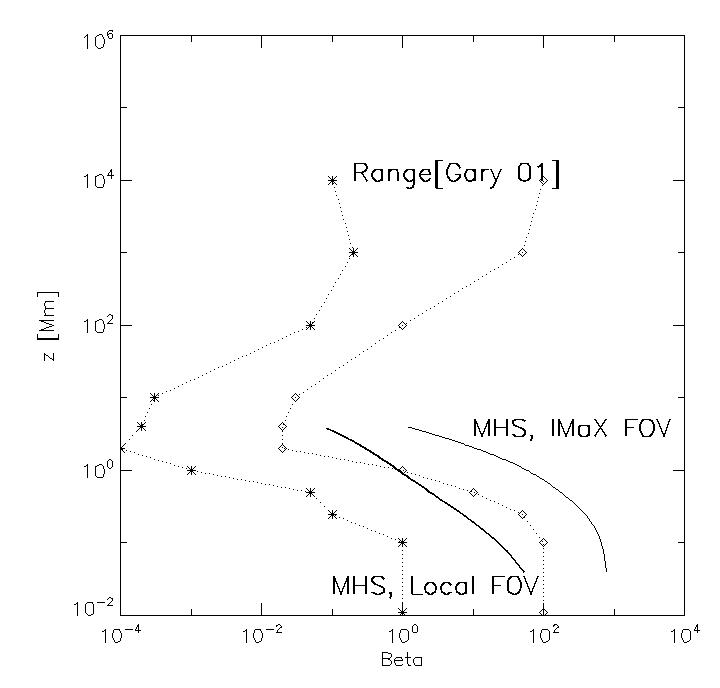}
\caption{Plasma $\beta$ in the solar atmosphere. The dotted lines are
taken from \cite{gary_01}. The thin solid line shows the (horizontally averaged)
plasma $\beta$ profile computed with our MHS-model for the full IMaX-FOV
and the thick solid line represents the same for the selected small area.}
\label{beta}
\end{figure}

In Fig. \ref{pic_pressure} a) we show the pressure
disturbance in the chromosphere at the height $z= 1 {\rm Mm}$
as calculated with the second term
$-\frac{1}{2 \mu_0} f(z) B_z^2$  on the right-hand-side
of Eq. (\ref{pressure}). This term obviously becomes
largest above regions with the highest photospheric field strength,
as seen in the large negative peaks. The total pressure has to be positive
of course and consequently a lower bound for the 1D-background pressure $P_0$
is given by Eq. (\ref{P0}). $P_0$ describes a 1D-equilibrium between
the gravity force and the vertical pressure gradient. One has to solve:
\begin{equation}
\frac{d P_0(z)}{d z} = -g \rho(z)
\end{equation}
for a constant gravity $g$.
Assuming an equation of state of the form
$
P_0=\rho R T
$
we get
\begin{equation}
\frac{d P_0(z)}{d z} = -\frac{g P_0(z)}{R T},
\label{def_P0}
\end{equation}
which leads to the well-known atmospheric exponential decay
$ \propto \exp(-\frac{z}{2H})$, with $H \approx 180 {\rm km }$.
for a constant
Temperature $T=T_0$, which is, however, not realistic for describing
structures reaching from the photosphere through the chromosphere into the
corona. Equation (\ref{def_P0}) can be (numerically) integrated for
any choice of a temperature profile $T(z)$, e.g., from 1D-models of
the solar atmosphere.
 Another alternative
is (because we computed already the 3D magnetic field
from Eq. \ref{lin_mhs} and  \ref{lin_mhs_2}) to prescribe
the average plasma $\beta(z)$ as a function of z, e.g. from the literature
\citep{gary_01}, leading to
\begin{equation}
P_0(z)=\frac{\beta(z) \, B^2_{\rm ave} }{2 \mu_0}
\end{equation}
where $B_{\rm ave}(z)$ is the horizontally averaged
$B_{z}(x,y,z)$. The allowed ranges for $\beta(z)$
are bounded from below, however, by Eq. \ref{P0}. A choice which
ensures a total positive pressure is obtained by using Eq. \ref{P0}
directly
\begin{equation}
P_0(z) = P_{\epsilon}(z)+  a \cdot \exp(-k z)
\cdot \frac{{\rm Max}(B_z)^2}{2 \mu_0}(z),
\label{max_p0}
\end{equation}
where $P_{\epsilon}(z)$ is the (prescribed) minimum value of
the total pressure at a given height. For $P_{\epsilon}(z)=0$
the total plasma pressure becomes zero at the maximum of $B_z$
and remains positive elsewhere. Taking
this into account we can calculate the full
average plasma $\beta$
(including the pressure disturbance) from Eq.
(\ref{pressure}), as shown in Fig.
\ref{beta} right-most-curve labeled {\it MHS, IMaX FOV} in
Fig. \ref{beta}. The limitations
from Eq. (\ref{P0}) and a magnetogram with some high peak values in
an otherwise weak field region cause values of plasma $\beta$ which
are too high and outside the range given by \cite{gary_01} (dotted curves).
We have to conclude, that the linear MHS-model cannot be applied to the
whole FOV of the SUNRISE magnetogram realistically. The reason is that through
Eq. (\ref{max_p0}) the 1-D background pressure and thereby
the maximum  pressure in weak field regions
is coupled with the highest values in the photospheric magnetogram,
which is not very realistic.

\subsection{Application to a small part of the FOV}
\begin{figure}
\includegraphics[width=0.5 \textwidth, bb= 80 30 670 500, clip = true]{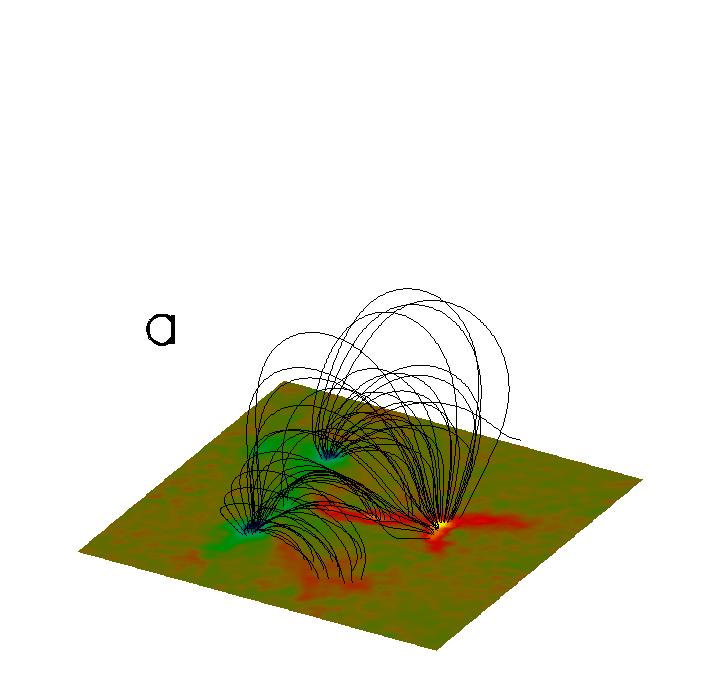}
\includegraphics[width=0.5 \textwidth, bb= 80 30 670 500, clip = true]{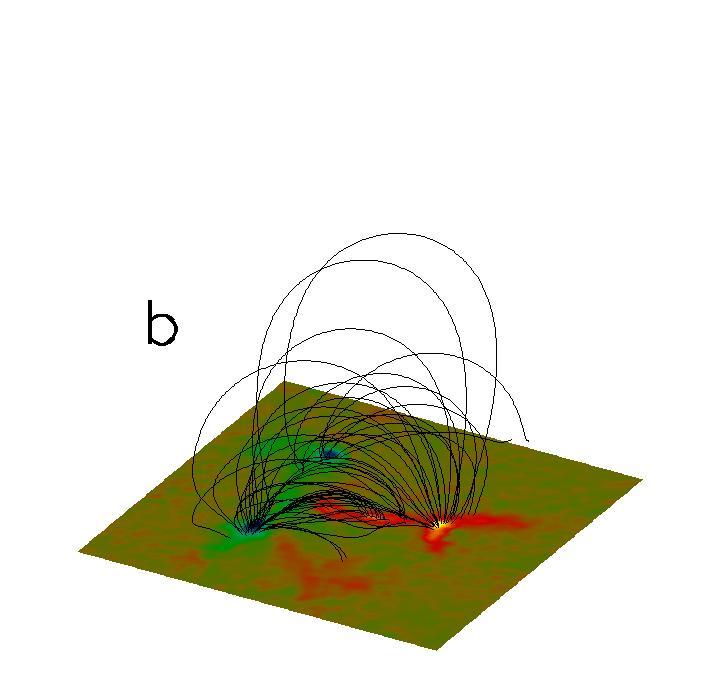}
\includegraphics[width=0.5 \textwidth, bb= 80 30 670 500, clip = true]{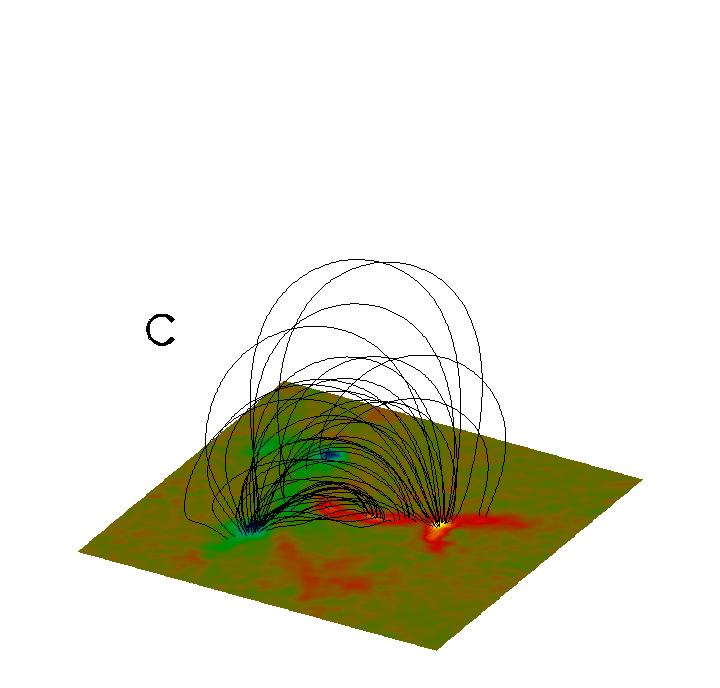}
\caption{Small field of view (rectangular box in fig. \ref{sunrise_mag_full}a).)
a: Potential field, b: Linear force-free field,
c: magneto-static field.}
\label{small_Bf}
\end{figure}

Due to the difficulties of applying the linear MHS-model to a full
magnetogram, we restrict our analysis in the following to the smaller sub-region
marked by the black rectangle in
Fig. \ref{sunrise_mag_full} a). Figure \ref{small_Bf} shows a few sample
field lines for a) a potential field model, b) a linear
force-free model with $\alpha=3$  and c) a magneto-static solution
$\alpha=3, \; a=0.5$.
In the linear force-free case the field lines become sheared compared
with the potential field and for some lines the connectivity changes.
The influence of a non-vanishing Lorentz force (but using the
same value of $\alpha$ as in the linear force-free case) has additional
effects, which seem, however, to be smaller. The maximum heights of
the loops are somewhat reduced and some additional field lines
change their connectivity, e.g. in the MHS-model no lines are
connected with the positive (red) flux region close to the front boundary.
Compared with the potential fields, the number of field lines connecting to
this region was already reduced in the linear-force-free model.

The pressure-disturbance in this smaller FOV is shown in the
center panel of Fig. \ref{pic_pressure}. The absence of strong peaks
in the photospheric field in this region leads to a much smoother
distribution
of the pressure disturbance. We use Eq. \ref{max_p0} to compute
the background pressure and in Fig. \ref{beta} the solid line marked
{\it MHS, local FOV} shows the averaged plasma $\beta$ as a function
of the height. At least in the
photosphere and chromosphere the plasma $\beta$ is within the limitation
given by the dashed lines from the literature \citep{gary_01}.
The true 3D plasma $\beta$ distribution is, however, not a function
of $z$ only, but varies significantly in the horizontal direction.
Fig. \ref{pic_pressure} c) shows the equi-contours
for $\beta=1.0$. As one can see the $\beta=1.0$ surface is by no means
plane-parallel, but strongly corrugated.
This behaviour impacts
methods for extrapolating force-free fields. Traditionally and for
numerical simplicity, one extrapolates from a plane parallel surface
(or the Sun's spherical surface) by assuming that the field above this
lower boundary of the computational domain is force-free. In reality,
however, the force-free domain is bounded  below by a corrugated
surface as well. This is also true for planned measurements of the
chromospheric magnetic field vector with Solar-C, so that magnetic field
extrapolation techniques bounded by non-plane-parallel surfaces should
be developed. In the non-force-free region between the photosphere and the
corrugated chromosphere, plasma pressure and gravity must be taken into
account.

\section{Discussion and Outlook}
\label{sec:outlook}
The linear-MHS approach used in this paper has 2 free parameters,
the linear-force-free
parameter $\alpha$ and the force-parameter $a$. Additionally one has to
prescribe, besides the vertical magnetic field component at the lower
boundary, also the height in the solar atmosphere, where the magnetic
field becomes approximately force-free, here $1/\kappa= 2 Mm$.
Applying these solutions to large-scale areas has its limitations.
These are, first of all,
the well-known limitations on $\alpha$, which these solutions
share with linear force-free configurations. Additionally the
pressure-gradient
(which compensates the Lorentz-force) is coupled to the vertical magnetic field.
As a consequence, the pressure disturbance, which is negative, becomes very
large above strong fields in the photosphere.
In order to maintain a positive total pressure,
the background plasma pressure must be so strong that the average
plasma $\beta$ becomes too high.
This limitation of the method has to do with the fact that
the two free parameters
$\alpha$ and $a$ have to be the same in the entire computational
domain. The limitations are similar as for linear force-free
fields, where one has only one free parameter $\alpha$, which
has to be globally constant. While linear force-free
fields cannot be used to model force-free configurations containing
strong current concentrations in part of the domain (leading to localized high
values of alpha), a similar restriction occurs here for the
linear magneto-static approach. Strong magnetic elements
in an otherwise weak field magnetogram cannot be modelled by this class
of MHS solutions.

These limitations do not occur, however, for application to
regions with a smaller field of view, because the assumption
that $\alpha$ and $a$ are constant is naturally more reasonable as smaller
the investigated domain is.
How should one proceed to derive global magneto-static
solutions? One possibility would be to compute the solutions discussed here
only locally (with different values of $\alpha$ and $a$ in different regions)
and to merge these configurations together. This will of course lead to
 solutions which are
not entirely self-consistent
and to inconsistencies at the boundaries between the
different regions. Another idea would be to use a numerical scheme, e.g.
an optimization approach  as suggested by
\cite{wiegelmann:etal06}, \cite{wiegelmann:etal07} to relax
these merged solutions
towards a self-consistent (nonlinear) MHS-equilibrium.
The methods developed  by \cite{wiegelmann:etal06} and
\cite{wiegelmann:etal07}
are both non-linear magneto static codes in cartesian and spherical
geometry, respectively. For the small scale features measured with Sunrise,
one naturally applies the cartesian version.
These codes require photospheric vector magnetograms
as input, which are not available for the investigated quiet Sun region,
because of the poor signal to noise ratio (for horizontal fields) in
weak field regions. Nonlinear approaches
(both force-free and magneto static) are well suited for dealing with
local strong enhancements (e.g. current concentrations and strong flux
elements). It is a weakness of any linear approach, that they cannot deal
with strong localized enhancements of any quantity.

To be able to carry out nonlinear magnetostatic
(or nonlinear force-free) extrapolations,
measurements of the horizontal photospheric
magnetic field, would be helpful.
During the re-flight of SUNRISE in 2013, high resolution vector
magnetograms
of active region(s) have been measured with IMaX and we plan to use
these measurements
for a self-consistent nonlinear magneto-static modelling in our future work.

\begin{acknowledgements}
The German contribution to {\it SUNRISE} is funded by the
Bundesministerium
 f\"{u}r Wirtschaft und Technologie through Deutsches Zentrum f\"{u}r Luft-
 und Raumfahrt e.V. (DLR), Grant No. 50~OU~0401, and by the Innovationsfonds of
 the President of the Max Planck Society (MPG).
 The Spanish contribution has
 been funded by the Spanish MICINN under projects ESP2006-13030-C06 and
 AYA2009-14105-C06 (including European FEDER funds). The HAO contribution was
 partly funded through NASA grant number NNX08AH38G.
 TN acknowledges support by the U.K.'s Science and Technology
 Facilities Council and would like to thank the MPS for its
 hospitality during a visit in December 2014.
 D.H.N. acknowledges financial support from GA \v{C}R under
grant number 13-24782S. The Astronomical Institute Ond\v{r}ejov is
supported by the project RVO:67985815.
\end{acknowledgements}
\bibliographystyle{aa}

\end{document}